\documentclass[11pt]{article}
\usepackage{graphicx}
\usepackage{amssymb}
\def\bddpmin{B^-\to D_s^- D^0\pi^0}
\def\dcbar{\,\, {\rule{0pt}{13pt}}^{\!\!\mbox{\tiny {(--)}}}\!\!\!\!\!\!}
\def\bddp{B^\pm\to D_s^\pm\dcbar D^0\pi^0}
\def\bddptit{B^\pm\to D_s^\pm\,\,
{\rule{0pt}{17pt}}^{\!\!\mbox{\scriptsize {(--)}}}\!\!\!\!\!\! D^0\pi^0}
\def\bddpabs{B^\pm\to D_s^\pm\,\,
{\rule{0pt}{11pt}}^{\!\!\mbox{\tiny {(--)}}}\!\!\!\!\!\! D^0\pi^0}

\def\be{\begin{equation}}
\def\ee{\end{equation}}
\def\bea{\begin{eqnarray}}
\def\eea{\end{eqnarray}}

\begin{document}
\thispagestyle{empty}
\begin{center}
\vspace*{-1cm}
{\Large \bf Looking for nonstandard CP violation \\
in $\bddptit$ decays}
\vspace*{1.2cm} \\
{ \sc L.N. Epele, D. G\'omez Dumm, A. Szynkman}
\vspace*{0.4cm} \\
{\em IFLP, Depto.\ de F\'{\i}sica, Universidad Nacional de La Plata, \\
C.C. 67, 1900 La Plata, Argentina}
\vspace*{0.7cm} \\
{\sc R. M\'endez--Galain}
\vspace*{0.4cm} \\
{\em Instituto de F\'{\i}sica, Facultad de Ingenier\'{\i}a, Univ. de la
Rep\'ublica, \\
C.C.\ 30, CP 11000 Montevideo, Uruguay}
\vspace*{1cm}
\begin{abstract}
We study the possibility of measuring nonstandard CP violation effects
through Dalitz plot analysis in \mbox{$\bddpabs$} decays. The accuracy in
the extraction of CP violating phases is analyzed by performing a
Monte Carlo simulation of the decays, and the magnitude of possible new
physics effects is discussed. It is found that this represents a hopeful
scenario for the search of new physics.
\end{abstract}
\vspace*{.7cm}
\end{center}

\vspace{1cm}

\section{Introduction}

The origin of CP violation in nature is presently one of the most
important open questions in particle physics. Indeed, the main goal of the
experiments devoted to the study of $B$ meson decays is either to confirm
the picture offered by the Standard Model (SM) or to provide evidences of
CP violation mechanisms originated from new physics. Among the various
CP-odd observables in $B$ physics, attention is mostly concentrated in the
``gold-plated'' channel $B \to J/\Psi\,K_S$. According to the SM picture,
from the analysis of a time-dependent CP asymmetry observed in these
decays it is possible to get a ``clean'' measurement of $\sin 2\beta$,
where $\beta$ is one of the angles of the so-called unitarity triangle
\cite{Nir94}. Recent measurements by BELLE and BaBar Collaborations,
together with previous results from Aleph, Opal and CDF, lead to the
(averaged) value $\sin 2\beta=0.734\pm 0.054$, which is in good agreement
with the constraints imposed by other measured CP-conserving and
CP-violating observables \cite{Nir02}.

In fact, the common belief is that the SM is nothing but an effective
manifestation of some underlying fundamental theory. In this way, all
tests of the standard mechanism of CP violation, as well as the
exploration of signatures of nonstandard physics, become relevant. One
important characteristic of the SM is that it includes only one source of
CP violation, namely a complex phase in the quark mixing matrix $V_{CKM}$.
In general, since overall phases of transition amplitudes are not
observable, one has to deal with interference effects in order to search
for measurable CP-violating quantities. Within the SM, there are some
specific processes in which the amplitude is either dominated by a single
contribution, or in which several contributions are significant, all of
them carrying the same weak phase. In these situations, weak SM phases are
unobservable, and asymmetries between CP conjugated processes are expected
to be vanishingly small. This offers an attractive window to search for
evidences of new physics, and is the main motivation for this work.

We show here that three body decays $B^+\to D_s^+ \bar D^0\pi^0$ and
$B^-\to D_s^- D^0\pi^0$ provide an interesting scenario to look for such
effects. For these processes, the main contributions to the decay
amplitude in the SM carry a common weak phase, therefore the measurement
of relative CP-violating phases, leading to an asymmetry between $B^+$ and
$B^-$ decays, would represent a signal of new physics. We discuss here the
possibility of performing these measurements by means of a Dalitz plot
(DP) fit analysis. In general, three body decays of mesons proceed through
intermediate resonant channels, and the DP fit analysis allows a direct
experimental access to the amplitudes and phases of the main
contributions~\cite{DP}. The usage of this technique for a clean
extraction of CP-odd phases has already been proposed in the
literature~\cite{Bed98} in relation with other CP-violating observables,
more precisely, to get clean measurements of the weak angle $\gamma$
within the SM. From the experimental point of view, the usage of charged
$B$ mesons has the advantage of avoiding flavor--tagging difficulties. In
addition, the processes $\bddp$ appear to be statistically favored, in
view of their relatively high branching ratios of about 1\%.

In order to evaluate the experimental perspectives, we perform a Monte
Carlo simulation of the actual processes, applying the DP fit technique to
evaluate the error in the extraction of possible CP-violating phases.
Then we perform a rough theoretical analysis, discussing the expected
magnitude of CP violation effects that could arise beyond the SM.
According to our results, the considered channels offer a promising
scenario to obtain a clear signature of new physics. In the worst case,
the lack of evidences would allow to improve the present bounds on the
parameters of the model under consideration.

The paper is organized as follows. In Sect.\ II we describe the general
framework, introducing the CP-violating observables. In Sect.\ III we
detail the DP fit procedure and present the results of our simulations.
Sect.\ IV is devoted to the theoretical discussion of new physics effects,
while in Sect.\ V we summarize our main results.

\section{CP-violating phases and Dalitz plot fit technique}

In this section we describe how the DP fit technique can be applied to
disentangle possible effects of new physics in $\bddp$ decays. In
principle, these processes are expected to proceed through various
intermediate resonances, as well as through a direct, nonresonant channel.
The total branching ratio will result from the interference of all these
contributions. The Dalitz plot maximum likelihood technique is a powerful
tool to get a clean disentanglement of the relevant contributing channels,
allowing to measure the ratios between the different partial amplitudes
{\it together with their relative phases}. This can be used to perform a
clean extraction of CP-violating phases, avoiding many theoretical
uncertainties.

Let us begin by summarizing the main steps of this procedure. More details
on these ideas can be found in Refs.\ \cite{Bed98}. In general, for a
given three body decay, in the DP fitting analysis of experimental data
one defines a fitting function ${\cal F} (m_1^2,m_2^2)$, where $m_1^2$ and
$m_2^2$ are the usual DP phase space variables. In our case this function
can be written as
\begin{equation}
{\cal F}_{B\to D_s D^0 \pi^0} (m^2_{1},m^2_{2}) =
| \Sigma_j \alpha_j e^{i\theta_j} A_j(m^2_1,m^2_2) |^2 \hskip 0.1 cm ,
\label{fit}
\end{equation}
where $m^2_1=(p_{\pi^0} + p_{\scriptscriptstyle D^0})^2$,
$m^2_2=(p_{\pi^0} + p_{\scriptscriptstyle D_s})^2$, $A_j$ are definite
functions corresponding to each partial channel, and $\alpha_j$ and
$\theta_j$ are real parameters that emerge as outputs from the fit. The
index $j$ labels the intermediate resonant channels, as well as the
nonresonant one. For the resonant channels, the main phase space
dependence of the functions $A_j$ is given by the Breit-Wigner (BW) shape
characterizing the resonances, together with definite angular functions
which depend on the spin of the corresponding resonant state (we come back
to this issues in the next section). The nonresonant decay amplitude is
assumed to be constant in most experimental analyses. This fitting
technique has proven to be successful to describe e.g.\ three body decays
of $D$ mesons \cite{D-decays}. In those analyses the phases $\theta_j$
have been extracted with combined statistical and systematic errors as
small as a few degrees, in experiments with a few thousands reconstructed
events.

In general, the phases $\theta_j$ can be written as the sum of a
``strong'' (CP-conserving) phase $\delta_j$ and a ``weak'' (CP-violating)
phase $\varphi_j$. These cannot be measured separately by a single fit.
Nevertheless, comparing the outputs from the CP-conjugated $B^+$ and $B^-$
decay experiments one can extract both phases $\delta_j$ and $\varphi_j$
simply from
\begin{eqnarray}
\delta_j & = & \frac12\, (\theta_j^+ + \theta_j^-)
\label{strong} \\
\varphi_j & = & \frac12\, (\theta_j^+ - \theta_j^-) \;,
\end{eqnarray}
where $\theta_j^+\,(\theta_j^-)$ stands for the phases measured from the
$B^+$ ($B^-$) decays. It is worth to notice that weak phases can be
extracted even in the limit where strong phases $\delta_j$ are vanishingly
small ---which is expected to be the case in many $B$ decays, owing to the
large $b$ quark mass. This represents a remarkable advantage with respect
to most proposals of measuring CP asymmetries in charged $B$ decays. In
general, in order to get a sizable asymmetry, one requires the presence of
strong FSI phases, which introduce a significant theoretical uncertainty.
In our case, however, strong phases are already supplied by the resonance
widths in the BW functions~\cite{Atw94}, and no theoretical estimation of
FSI phases is needed. Moreover, the latter can be independently obtained
from the fit by means of Eq.~(\ref{strong}).

It is important to notice that for the fitting procedure to apply, it is
necessary that the decay amplitude receives contributions from at least
two intermediate channels carrying different CP violating phases. Indeed,
an overall phase is physically meaningless, and the DP fit only allows the
measurement of relative phases between different channels.

Let us now analyze the case of the decays $\bddp$ in the framework of a
theory including physics beyond the SM. For each intermediate channel, it
is natural to assume that new physics occurs at a relatively high energy
scale, therefore its effects can be decoupled from the resonance BW
functions, the angular functions, and other possible form factors in
$A_j(m_1^2,m_2^2)$ arising from low-energy hadronic interactions.
Accordingly, the fitting function ${\cal F}$ will be still of the form
proposed in Eq.\ (\ref{fit}), with the same functions $A_j$, and now the
complex weights $\alpha_j\, e^{i\theta_j}$ will include the effects of new
physics:
\begin{equation}
\alpha_j^{\pm(exp)}\; e^{i \theta_j^{\pm(exp)}} =
\alpha^{SM}_j\;e^{i(\delta^{SM}_j \pm \varphi^{SM}_j)} +
\alpha^{NP}_j\;e^{i(\delta^{NP}_j \pm \varphi^{NP}_j)}\;.
\label{ampgen}
\end{equation}
Here the index $(exp)$ refers to the experimentally measurable quantities
(outputs of the DP fit), whereas $SM$ and $NP$ denote Standard Model and
new physics contributions respectively. The $\pm$ signs correspond to
decays of $B^\pm$ mesons.

Within the SM, the short-distance effective Hamiltonian relevant for the
decays $\bddp$ has been studied in detail \cite{Buch96}, including the
effects of strong and electroweak penguin operators. The situation is
similar as in the ``gold-plated'' channel $B\to J/\Psi K_S$, in the sense
that the main contributions (both tree and penguin) to the effective
operators carry the same weak phase. In this way, this phase is expected
to factorize, being common to all (resonant and nonresonant) channels
contributing to the decay. One could thus conventionally set
$\varphi_j^{SM}=0$ for all $j$, and no CP asymmetry should be observed
between $B^+$ and $B^-$ decay patterns in absence of new
physics\footnote{Within the SM, one expects in fact a tiny CP asymmetry in
the decay rates. This has been analyzed in Ref.~\cite{Gir01} for the
process $B^-\to D^0 D_s^-$, where the effect is found to be about 0.2\%.}.
On the contrary, if new physics is present, the situation may be
different. To simplify the analysis, let us assume that only two
intermediate channels contribute, namely those mediated by resonances
\mbox{$\dcbar D^{\ast 0}$} and $D_s^{\ast\pm}$ ---say channels 1 and 2,
respectively. This is a natural assumption, since in fact they are
expected to largely dominate the decay (in any case, if other intermediate
channels were shown to provide significant contributions, the procedure we
describe here can still be followed on the same grounds). In this
two-channel case, only the relative phases between both contributions 1
and 2 can be measured. From Eq.\ (\ref{ampgen}), one has
\begin{equation}
\theta^{\pm(exp)} \equiv \theta_1^{\pm(exp)}-\theta_2^{\pm(exp)}
= \arg\; \left[  \frac{\alpha^{SM}_1 +
\alpha^{NP}_1\;e^{ i (\delta_1^{NP}-\delta_1^{SM}\pm \varphi_1^{NP}) }}
{\alpha^{SM}_2 + \alpha^{NP}_2\;e^{ i (\delta_2^{NP}-
\delta_2^{SM}\pm \varphi_2^{NP}) }} \right]
+ \delta_1^{SM} - \delta_2^{SM}\;.
\label{diff}
\end{equation}

The theoretical framework can be simplified by introducing some natural
assumptions. First, it is reasonable to think that the resonance
hadronization and decay processes ---which are governed by strong
interactions in the nonperturbative regime--- can be disentangled from the
effects of new physics, the latter taking place at a high energy scale. In
addition, it is usual to assume that strong FSI are the main source for
strong phases $\delta_j$, since high energy contributions to CP-conserving
phases arising from absorptive parts of QCD and electroweak loop diagrams
are shown to be suppressed \cite{Ger91}. In this way, for each resonant
channel strong phases should factorize out, i.e.\ $\delta^{NP}_j =
{\delta^{SM}_j} \equiv \delta_j$. On the other hand, in most scenarios of
new physics ---as well as in the SM \mbox{itself---,} CP-violating phases
are essentially determined by the flavor content of the quarks entering
the diagrams that dominate the $b$ quark decay. If this is the case, since
both resonant states $D^{\ast 0} D_s$ and $D^0 D_s^\ast$ have the same
quark content, one expects that the new CP-violating phases obey
$\varphi^{NP}_1 = \varphi^{NP}_2 \equiv \varphi^{NP}$, remaining constant
along the phase space. Once these assumptions have been taken into
account, the measurable complex weights in Eq.\ (\ref{ampgen}) can be
written as
\begin{equation}
\alpha_j^{(exp)}\; e^{i \theta_j^{\pm(exp)}} =
 ( \alpha^{SM}_j +
\alpha^{NP}_j\;e^{ \pm i \varphi^{NP})}) \; e^{i \delta_j}
\;,\quad j=1,\,2\; .
\label{ampgenred}
\end{equation}
We have dropped here the $\pm$ signs in $\alpha_j^{(exp)}$, since the
assumption $\delta_j^{NP}=\delta_j^{SM}$ implies $\alpha_j^{+(exp)} =
\alpha_j^{-(exp)}$ \footnote{In principle, this last relation could be
experimentally checked through the comparison between the fits for $B^+$
and $B^-$ decays, providing a consistency test for our assumptions on the
strong phases.}. The expression for the relative phases
$\theta^{\pm(exp)}$ in Eq.\ (\ref{diff}) simplifies now to
\begin{equation}
\theta^{\pm(exp)} = \arg\; \left ( \frac{\alpha^{SM}_1 +
\alpha^{NP}_1\;e^{ \pm i \varphi^{NP}}}{\alpha^{SM}_2 + \alpha^{NP}_2\;e^{
\pm i \varphi^{NP}}} \right ) + \delta_1 - \delta_2\;.
\end{equation}

As stated, we are interested in the difference between the relative phases
for the CP-conjugated decays $B^+$ and $B^-$, which is an observable of CP
violation. This is given by
\begin{equation}
\Delta \theta^{(exp)} \equiv \theta^{+(exp)} - \theta^{-(exp)} = \arg \left
( \frac{\alpha^{SM}_1 + \alpha^{NP}_1\;e^{+ i \varphi^{NP}}}{\alpha^{SM}_2
+ \alpha^{NP}_2\;e^{+ i \varphi^{NP}}} \right ) - \arg \left (
\frac{\alpha^{SM}_1 + \alpha^{NP}_1\;e^{- i \varphi^{NP}}}{\alpha^{SM}_2 +
\alpha^{NP}_2\;e^{- i \varphi^{NP}}} \right )\;.
\label{CPV}
\end{equation}
Finally, assuming that new physics contributions are small when compared
to SM amplitudes, i.e., $\alpha^{NP}_j \ll \alpha^{SM}_j$, we end up with
\begin{equation}
\Delta \theta^{(exp)} \simeq 2 \, \sin \varphi^{NP} \left(
\frac{\alpha^{NP}_1}{\alpha^{SM}_1} - \frac{\alpha^{NP}_2}{\alpha^{SM}_2}
\right )\,.
\label{relphase}
\end{equation}
Notice that this quantity is independent of any CP-conserving phase. It
only depends on the (in principle, unknown) real amplitudes
$\alpha_j^{SM}$ and $\alpha_j^{NP}$, and on the (also unknown)
CP-violating phase $\varphi^{NP}$.

As we have discussed above, $\Delta \theta^{(exp)}$ vanishes in the
absence of new physics. This is a convenient situation for the search of
clean effects of physics beyond the SM, provided that the factors in the
r.h.s.\ of Eq.\ (\ref{relphase}) are large enough to allow a clear
experimental signature. The perspectives in this sense are addressed in
the next sections.

\section{Experimental perspectives}

In this section we make an estimate of the precision that may be
reached in the measurement of the phase difference $\Delta \theta^{(exp)}$
in $\bddp$ decays. This will indicate, according to the result in Eq.\
(\ref{relphase}), the minimum size of new physics contributions to the
decay amplitudes needed to yield a distinguishable experimental
signal.

One important reason for which the channels considered here deserve
special attention is their relatively high statistics. Since the branching
ratios for $\bddp$ are as large as $\sim$ 1\%, after a couple of years of
full run of LHCb, and assuming a 20\% reconstruction efficiency, one
should end up with some $10^5$ reconstructed events in each $B^+$ and
$B^-$ Dalitz plots. This is a large number, taking into account that DP
fits performed for $D$ meson decays with much less events have led to the
measurement of relative phases with statistical errors of just a few
degrees \cite{DP}. However, the processes considered here are very
different from those. Indeed, even if such a large number of events will
certainly give a very precise measurement of the branching fractions for
each partial channel ---i.e., the quantities $\alpha_j^{(exp)}$---, a
precise measurement of phases requires not only large statistics but also
a large interference region between the different intermediate channels.
It is not obvious that this will be the case for $\bddp$ decays, since the
involved resonances are very narrow, their widths laying below 1
MeV~\cite{PDG}.

In order to evaluate the actual experimental feasibility of our proposal,
we have carried out a Monte Carlo simulation of the decays. Our goal is to
generate $10^5$ events in the Dalitz plot, and then to perform a Dalitz
plot fit analysis in order to determine if the phases can be successfully
extracted with a small statistical error. Clearly, this simulation does
not account for the details concerning the detectors. The possible impact
of systematic errors will be discussed below.

We have generated $10^5$ events using a decay amplitude of the form in
Eq.~(\ref{fit}). As a first guess, we include in the decay only three
channels, namely those mediated by the resonances $\dcbar D^{\ast 0}$ and
${D_s^\ast}^\pm$, and the direct nonresonant decay $B^\pm\to (D_s^\pm
\dcbar D^0\pi^0)_{NR}$. The form of the functions $A_j$ for the resonances
$j=1,2$ is \cite{DP}
\begin{equation}
A_j = BW_{j}(m_{j}^2)\, (\vec p_B \cdot \vec p_\pi)\, F_j(m^2_j)\;,
\label{ampl}
\end{equation}
where the invariant masses $m_j^2$ are defined as in Eq.\ (\ref{fit}),
$F_j(m_j^2)$ is a form factor, and $BW_j(s)$ is the Breit-Wigner function
\begin{equation}
BW_j(s) = \frac{1}{m_{R_j}^2 - s - i m_{R_j} \Gamma_{R_j}(s)}\;,
\label{BW}
\end{equation}
$m_{R_j}$ being the resonance masses ($R_1=\dcbar D^{\ast 0}$,
$R_2={D_s^\ast}^\pm$). For each $j$, the $B$ and $\pi$ meson three-momenta
in Eq.\ (\ref{ampl}) are evaluated in the rest frame of the corresponding
intermediate resonance.

We have taken the usual expressions~\cite{cg-3pi} for the form
factors\footnote{The actual shape of the form factors in $B$ decays is in
general unknown. We have considered expressions similar to those used for
$D$ decays, finding that their incidence is not relevant to the discussion
in this work.} and for the momentum-dependent width $\Gamma_{R_j}(s)$. The
latter is given by
\begin{equation}
\Gamma_{R_j}(s) = \Gamma_{R_j}\; \frac{m_{R_j}}{\sqrt{s}}
\left|\frac{\vec p\,(s)}{\vec p\,(m_{R_j}^2)} \right|^3\;,
\end{equation}
where $\Gamma_{R_j}$ is the on-shell resonance width, and $\vec p\,(q^2)$
stands for the three-momentum of the resonance decay products when the
resonance mass is $\sqrt{q^2}$. The shape of the nonresonant decay
amplitude, which is in general unknown~\cite{Bed97}, has been taken ---as
it is usually done--- as a constant function. In any case, as it is
discussed below, this assumption has a negligible impact on our results.

In order to carry out the generation of events, we need to introduce as
input data the values for the physical quantities $\alpha_j$, $\theta_j$
and the resonance widths. The expected relative weights $\alpha_j$ of the
two resonant channels can be obtained from the known branching ratios
$BR(B^- \to D^{\ast 0} D_s^-)$, $BR(D^{\ast 0} \to D^0 \pi^0)$, $BR(B^-
\to {D_s^\ast}^- D^0)$ and $BR({D_s^\ast}^- \to D_s^- \pi^0)$. We
have~\cite{PDG}
\begin{equation}
\frac{\alpha_1}{\alpha_2} = \sqrt{\frac{BR(B^- \to D^{\ast 0}
D_s^-)\,\times\, BR(D^{\ast 0} \to D^0 \pi^0)} {BR(B^- \to {D_s^\ast}^-
D^0)\,\times\, BR({D_s^\ast}^- \to D_s^- \pi^0)}}\; \sim \; 4\; .
\end{equation}
On the other hand, the nonresonant decay amplitude is uncertain; it is
just expected to be smaller than the resonant channels. Taking into
account that only the relative values between the three coefficients
$\alpha_j$ have a physical meaning in the simulation, we have taken
$\alpha_1 = 1$ and $\alpha_2 = 0.25$, while for $\alpha_{NR}$ we have
considered different values, ranging from 0 to 0.1. Concerning the phases
$\theta_j$, one expects CP-conserving parts to be relatively small,
whereas the CP-violating SM phase is essentially the same for all
amplitudes and can be factorized out. Thus, assuming that SM contributions
dominate, it is reasonable to choose all phases $\theta_j$ in our
numerical simulation to be small numbers. In fact, it will be seen that
this assumption is not relevant to our conclusions.

Finally, other relevant inputs in our simulation are the on-shell widths
$\Gamma_{R_j}$ of both resonances: events coming from a narrow resonance
should be concentrated in a given region of the plot, hence the
interference region between both resonances is expected to be relatively
small. Present measurements of $\dcbar D^{\ast 0}$ and ${D_s^\ast}^\pm$
widths are not conclusive, giving in both cases only upper bounds of about
2 MeV. For our simulations, we have chosen to consider values ranging from
0.01 to 1 MeV. The recent measurement of the ${D^\ast}^+$ width, which is
found to be around 0.1 MeV \cite{Cin01}, can be thought as a hint of the
expected orders of magnitude.

\begin{figure}[htb]
\centerline{
   \includegraphics[height=5truecm]{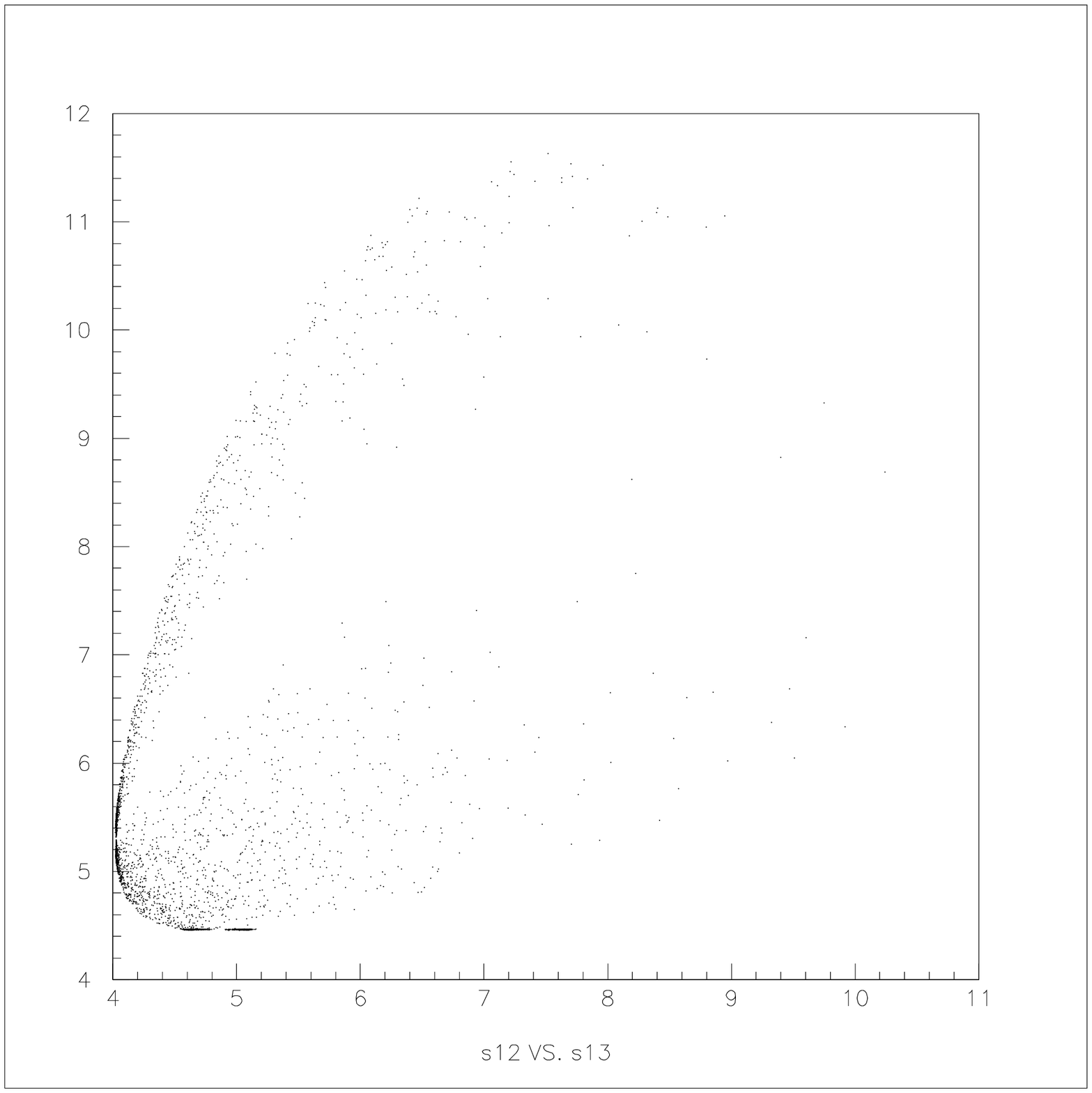}
   }
\caption{Dalitz plot for the $\bddpmin$ decay}
\end{figure}

As an example, we show in Fig.\ 1 the Dalitz plot generated with
$\alpha_{NR}=0.1$, $\theta_1=0$, $\theta_2=20^\circ$,
$\theta_{NR}=10^\circ$, and equal widths of 1 MeV for both resonances
(besides the already given values of $\alpha_1=1$ and $\alpha_2=0.25$).
One observes that, even if both resonances are quite narrow, the events
appear to be spread out in a large region of the plot. This is the
consequence of a purely kinematic effect, due to the fact that both
resonances are located very close to the threshold of the phase space.
This effect compensates the narrow width suppression, and brings a good
hope to extract the relative phases successfully.

After carrying out this simulation of the decay, we have performed a fit
of the data according to the fitting function given in Eq.\ (\ref{fit}),
where now the coefficients $\alpha_j$ and $\theta_j$ are left as free
parameters. In fact, as explained above, the fit provides only {\em
relative} values for both amplitudes and phases \cite{DP}, therefore we
have kept fixed the reference values $\alpha_1=1$ and $\theta_1=0$. The
result of the fit is given in Table \ref{tab1}. The method allows to
extract the phase $\theta_2$ with a statistical error as small as
$1.4^\circ$.

\begin{table}[htb]
\begin{center}
\hfill \\
\begin{tabular}{c||c|c}
channel  & $\alpha_j$ & $\theta_j$  \\
    \hline
$D^{\ast 0} D_s$ & fixed & fixed \\
$D^0 D_s^\ast$ & $0.2514\pm 0.0017$ & $(20.7\pm 1.4)^\circ$ \\
nonresonant & $0.1007\pm 0.0020$ & $(9.1\pm 1.2)^\circ$
  \end{tabular}
  \caption[]{Fitting results of the Monte Carlo sample. The events have
been generated with $\alpha_2=0.25$,
$\theta_2=20^\circ$, $\alpha_{NR}=0.1$, $\theta_{NR}=10^\circ$, and
$\Gamma_{D^{\ast 0}}=\Gamma_{D_s^\ast} = 1$ MeV.}
\label{tab1}
\end{center}
\end{table}

We have performed a systematic study of the results of the fit allowing
reasonable ranges of variation for the unknown quantities used to generate
the Monte Carlo sample, namely the resonance widths, the weight
$\alpha_{NR}$ and the relative phases $\theta_j$. As a first outcome of
this analysis, it is found that the statistical errors are independent of
the initial values of the phases. Secondly, the errors for both the
extracted amplitude and phase of the $D_s^\ast$ mediated decay (channel 2)
are independent of the weight $\alpha_{NR}$ of the nonresonant channel,
{\it even in the limit $\alpha_{NR}=0$}. This shows that the interference
between the two resonant channels is not mediated by the nonresonant one,
but arises from the above mentioned spread out of the events corresponding
to resonance-mediated decays. Finally, as expected, it is found that the
errors in the extracted weights $\alpha_j$ are independent of the
resonance widths; on the contrary, the values of the widths do affect the
quantity we are interested in, i.e.\ the error in the extracted relative
phase $\theta_2 - \theta_1$. This dependence is illustrated by the results
in Table~\ref{tab2}, where we have considered several simulations in which
the amplitudes and phases $\alpha_j$, $\theta_j$ have been taken as in the
previously described example. We quote in the Table the errors obtained in
the extraction of $\theta_2 - \theta_1$ for different values of
$\Gamma_{D^{\ast 0}}$ and $\Gamma_{D_s^\ast}$. In the first five rows of
the Table we have assumed equal ${D^\ast}^0$ and $D_s^\ast$ widths, while
in the last row we have taken $\Gamma_{D^{\ast 0}} = 100$ KeV,
$\Gamma_{D_s^\ast} = 10$ KeV (in fact, a relative suppression of the
$D_s^\ast$ width could be expected since the strong decay
${D_s^\ast}^\pm\to D_s^\pm \pi^0$ violates isospin). We see here that for
a width as narrow as 10 KeV the phase difference can still be extracted
with relatively low statistical error.

\begin{table}[htb]
\begin{center}
  \begin{tabular}{c c||c}
$\Gamma_{D^{\ast 0}}\,;\Gamma_{D_s^\ast}$ & \hspace{-.5cm} (MeV) & Error \\
    \hline
1 & & $1.4^\circ$ \\
0.5 & & $1.5^\circ$ \\
0.1 & & $1.7^\circ$ \\
0.05 & & $2.3^\circ$ \\
0.02 & & $5.1^\circ$ \\
$\;\;0.1\; ;\; 0.01$ & & $4.0^\circ$
  \end{tabular}
  \caption[]{Errors in the extracted value of $\theta_2-\theta_1$, for
different values of resonance widths. Input amplitudes and phases for
the event generation are same as in Table I.} \label{tab2}
\end{center}
\end{table}

Before ending this section let us say a few words about systematic
(experimental) errors in the extraction of phases. The evaluation of these
errors is in general a quite difficult task. In order to carry out the
complete analysis, one should perform a full numerical simulation of the
experiment including the detector, which is out of the scope of this
paper. Nevertheless, in order to have an estimate we can take into account
the results from recent DP analyses \cite{cg-3pi,cg-kpipi}. The latter
suggest that the systematic error in the measurement of phases for
intermediate channels with large branching fractions should not be above a
few degrees, i.e.\ of the same order of those quoted in Table~\ref{tab2}.

\section{Expected size of new physics effects}

Let us now turn back to Eq.\ (\ref{relphase}) and analyze the theoretical
expectations for the size of $\Delta\theta^{(exp)}$ in the context of a
theory beyond the SM, in order to evaluate if this observable has
potential chances to provide experimental evidences of new physics. To
carry out the theoretical analysis we take into account the low-energy
effective Hamiltonian relevant for the processes under consideration,
including QCD corrections at the leading order. Then, to deal with
long-range matrix elements, we use the simple factorization
approach~\cite{bau87}, which should be adequate to estimate the
significance of the new contributions \cite{Kim02}.

In view of the large hadronic uncertainties and the usual amount of
freedom to fix new physics parameters, we do not intend to perform an
accurate calculation of possible nonstandard contributions to the $\bddp$
decay amplitude. Just as an illustrative example, we consider the rather
representative framework of multihiggs models, showing that the situation
becomes quite promising if nonstandard contributions to penguin diagrams
are comparable to those arising from SM physics.

Our theoretical analysis is based on the $\Delta B=1$ effective
Hamiltonian \cite{Buch96,AliGre98}
\begin{eqnarray}
{\cal H}_{\rm eff} = \frac{G_F}{\sqrt 2} \left\{ V_{cb} V^*_{cs} (C_1 O_1 +
C_2 O_2) - V_{tb} V^*_{ts} (\sum_{i=3}^7 C_i O_i) \right\}\,,
\label{hamilt}
\end{eqnarray}
where $C_i$ are Wilson coefficients evaluated at a renormalization
scale $\mu\approx m_b$, and $O_i$ are local operators,
\begin{equation}
\begin{array}{rclrcl}
O_1 & = & (\bar c_\alpha \, b_\alpha)_{V-A}
(\bar s_\beta \, c_\beta)_{V-A} \hspace*{3.cm} &
O_2 & = & (\bar c_\beta \, b_\alpha)_{V-A}
(\bar s_\alpha \, c_\beta)_{V-A} \\
O_3 & = & (\bar s_\alpha \, b_\alpha)_{V-A}
\sum_{q'} (\bar q'_\beta \, q'_\beta)_{V-A} &
O_4 & = & (\bar s_\beta \, b_\alpha)_{V-A}
\sum_{q'} (\bar q'_\alpha \, q'_\beta)_{V-A} \\
O_5 & = & (\bar s_\alpha \, b_\alpha)_{V-A}
\sum_{q'} (\bar q'_\beta \, q'_\beta)_{V+A} &
O_6 & = & (\bar s_\beta \, b_\alpha)_{V-A}
\sum_{q'} (\bar q'_\alpha \, q'_\beta)_{V+A} \\
O_7 & = & (g_s/8\pi^2) m_b\,\bar s_\alpha\,\sigma^{\mu\nu}
\,(1+\gamma_5)\, T^a_{\alpha\beta}\, b_\beta\, G^a_{\mu\nu}\;. &
\end{array}
\label{oper}
\end{equation}
Here $V\pm A$ refers to the Lorentz structure $\gamma_\mu(1\pm\gamma_5)$,
$\alpha$ and $\beta$ stand for $SU(3)$ color indices, $T^a_{\alpha\beta}$
are generators of $SU(3)$ color transformations and $G^a_{\mu\nu}$ denotes
the gluonic field strength tensor. Contributions from electroweak penguins
will not be taken into account, therefore these operators have not been
included in (\ref{hamilt}). We will also neglect the effect of the
electromagnetic dipole operator. Within the SM, the coefficients $C_i$ can
be calculated at the scale $m_W$, and then evolved to $\mu\approx m_b$
through the renormalization group equations \cite{Buch96}. The $V_{CKM}$
factors corresponding to each operator have been explicitly separated in
(\ref{hamilt}), so that with good approximation the coefficients $C_i$ in
the SM can be assumed to be real numbers\footnote{In fact, they carry
small CP-violating and CP-conserving phases, coming from
Cabibbo-suppressed contributions and absorptive parts of loop diagrams
respectively.}. Moreover, in view of the unitarity of the $V_{CKM}$
matrix, one has $V_{tb} V^*_{ts} = - V_{cb} V^*_{cs} - V_{ub}
V^*_{us}\simeq - V_{cb} V^*_{cs}$, where the correction due to the $V_{ub}
V^*_{us}$ term is about 5\%. In this way, for the case under
consideration, the CP-violating phase carried by the penguin contributions
in the SM is approximately the same as that coming from the tree operators
$C_1$ and $C_2$, and will factorize out for the decay amplitudes of
interest (the contribution of the $V_{ub} V^*_{us}$ term to the full
amplitude will be below 0.5\% if, as expected, the total penguin amplitude
does not exceed 10\% of the tree piece). In a given extension of the SM,
however, the coefficients $C_i$ will carry in general nonvanishing
CP-violating phases $\varphi_i$, allowing for the interference effects
discussed in the previous sections.

In general, in a theory including physics beyond the SM, one expects that
the new particles can be integrated out at the $m_W$ scale, leading to new
contributions to the coefficients $C_i(m_W)$. However, since the new
particles have been integrated out, the running of the coefficients down
to $\mu\approx m_b$ proceeds just as in the SM~\cite{Xia01}. This running
of SM coefficients has been analyzed in detail in
Refs.~\cite{Buch96,Bur98} and will not be repeated here.

In the evaluation of the amplitudes $\langle VP|{\cal H}_{\rm
eff}|B\rangle$, the scale and renormalization scheme dependence introduced
by the coefficients $C_i$ should be compensated by that of the matrix
elements of the quark operators $O_i$ between the hadronic states.
However, as stated above, to evaluate these quantities we will use the
factorization ansatz, and in this approach the matrix elements are written
in terms of decay constants and form factors, which are both scale and
renormalization scheme independent. In order to achieve the required
cancellation, it is possible \cite{AliGre98} to calculate the one-loop
corrections to the partonic matrix elements $\langle s\bar c
c|O_i|b\rangle$, and to define new effective coefficients $C_i^{\rm eff}$
such that the one-loop quark-level matrix elements read
\begin{equation}
\langle s\bar c c|{\cal H}_{\rm eff}|b\rangle = \sum_{i=1}^6 C_i^{\rm eff}\langle
s\bar c c|O_i|b\rangle^{tree}\;.
\end{equation}
At NLO these effective coefficients will be given by the original $C_i$
plus QCD corrections,
\begin{equation}
C_i^{\rm eff} = C_i(\mu) + \frac{\alpha_s}{4\pi}\, \sum_{j=1}^7 K_{ij}(\mu)\,
C_j(\mu)\;.
\label{efect}
\end{equation}
The analytic expressions for the functions $K_{ij}$ can be found in
Refs.~\cite{AliGre98,Ali98,Che99}. It can be shown that now the effective
coefficients $C_i^{\rm eff}$ are scale and scheme independent, as well as
gauge invariant and infrared safe~\cite{Cheli99}. An important point is that
the corrections introduced in Eq.~(\ref{efect}) involve the coefficient
$C_7$, which can receive important contributions coming from nonstandard
physics, as occurs e.g.\ in the case of two-Higgs-doublet models
\cite{Xia01}. Even if the operator $O_7$ does not contribute directly to
the $B\to VP$ decay amplitudes in the factorization approach, the
combination in (\ref{efect}) implies that the new physics corrections to
$C_7$ are translated to other effective coefficients $C_i^{\rm eff}$ and
thus to the decay amplitude.

The previous analysis can be now applied to the decays of our interest,
namely the resonant processes $B^-\to D_s^{\ast -} D^0$; $D_s^{\ast -} \to
D_s^- \pi^0$ and $B^-\to D^{\ast 0} D_s^-$; $D^{\ast 0} \to D^0 \pi^0$
that dominate the three body decay $B^-\to D^0 D_s^-\pi^0$. In the
described framework, the relevant two-body amplitudes $\langle D_s^{\ast
-} D^0 |{\cal H}_{\rm eff}|B^-\rangle$ and $\langle D^{\ast 0} D_s^-
|{\cal H}_{\rm eff}|B^-\rangle$ will be given by
\begin{eqnarray}
\langle V P |{\cal H}_{\rm eff}|B^-\rangle & = &
\frac{G_F}{\sqrt{2}}\, V_{cb} V_{cs}^\ast
\sum_{i=1}^6 C_i^{\rm eff} \langle V P |O_i| B^-\rangle_{FA}
\nonumber \\
& = & \frac{G_F}{\sqrt{2}}\, V_{cb} V_{cs}^\ast\,
\tilde a(B^-\to VP)\, X^{(B^-\to VP)}\;,
\label{ampfa}
\end{eqnarray}
where the subindex $FA$ denotes that the matrix element is evaluated
within the factorization approximation. The factor $\tilde a(B^-\to VP)$
includes the effective coefficients $C_i^{\rm eff}$, whereas $X^{(B^-\to
VP)}$ contains the form factors related to the factorized amplitudes. For
the processes under consideration one has \cite{cheng99,Kim01}
\begin{eqnarray}
\tilde a(B^-\to D^{\ast 0} D_s^-) & = & a_1 + a_4 - 2\, a_6
\frac{m_{D_s}^2}{(m_b+m_c)(m_s+m_c)} \nonumber \\
\tilde a(B^-\to D_s^{\ast -} D^0) & = & a_1 + a_4 \;,
\label{a2}
\end{eqnarray}
where the coefficients $a_i$ are defined as $a_i \equiv C_i^{\rm eff} +
C_{i+1}^{\rm eff}/(N_c^{\rm eff})_i$ for $i=1$, and $a_i \equiv
C_i^{\rm eff} + C_{i-1}^{\rm eff}/(N_c^{\rm eff})_i$ for $i=4,6$. The
effective parameters $(N_c^{\rm eff})_i$ in these expressions account for
the uncertainties introduced when calculating the matrix elements of the
effective operators between hadron states \cite{AliGre98,Ali98,Che99}. The
factors $X^{(B^-\to VP)}$ are given by
\begin{eqnarray}
X^{(B^-\to D^{\ast 0} D_s^-)} & = & 2\,f_{D_s}\, m_{D^{\ast
0}}\,A_0^{BD^\ast}(m_{D_s}^2)\,
(\varepsilon_{D^{\ast 0}}^\ast\cdot P_B)
\nonumber \\
X^{(B^-\to D_s^{\ast -} D^0)} & = & 2\,f_{D_s^\ast}\,
m_{D_s^\ast}\,F_1^{BD}(m_{D_s^\ast}^2)\,
(\varepsilon_{D_s^\ast}^\ast\cdot P_B)\;,
\label{x2}
\end{eqnarray}
where $\varepsilon_V$ are the corresponding $V$ meson polarizations, $P_B$
is the $B$ four-momentum, and the expressions include decay constants and
form factors that can be estimated in specific models. In fact, Eqs.
(\ref{x2}) have been quoted only for completeness, since the factors
$X^{(B^-\to VP)}$ cancel out in our estimation for $\Delta\theta^{(exp)}$.
This can be seen by noticing that the expression for
$\Delta\theta^{(exp)}$ in (\ref{relphase}) involves ratios between SM and
new physics amplitudes. According to previous assumptions, the effects of
new physics are only present in the effective coefficients $C_i^{\rm eff}$
---or, equivalently, $\tilde a(B\to VP)$---, therefore any global factor
will cancel. One has in this way
\begin{eqnarray}
\frac{\alpha^{NP}_1 \, e^{-i\varphi^{NP}}}{\alpha^{SM}_1} & = &
\frac{\langle D^{\ast 0} D_s^-
|{\cal H}_{\rm eff}|B^-\rangle^{NP}}{
\langle D^{\ast 0} D_s^-|{\cal H}_{\rm eff}|B^-\rangle^{SM}}\;\
\simeq \;\ \frac{(a_1 + a_4 - 2\, r\, a_6)^{NP}}{(a_1 + a_4 - 2\,r\, a_6)^{SM}}
\nonumber \\
\frac{\alpha^{NP}_2 \, e^{-i\varphi^{NP}}}{\alpha^{SM}_2} & = &
\frac{\langle D_s^{\ast -} D^0
|{\cal H}_{\rm eff}|B^-\rangle^{NP}}{
\langle D_s^{\ast -} D^0|{\cal H}_{\rm eff}|B^-\rangle^{SM}}\;\
\simeq \;\ \frac{(a_1 + a_4)^{NP}}{(a_1 + a_4)^{SM}} \;,
\label{acoc}
\end{eqnarray}
where $r$ stand for the mass ratio $m_{D_s}^2/[(m_b+m_c)(m_s+m_c)]$, and
---as in the previous sections--- we have assigned labels 1
and 2 to the channels mediated by the resonances $D^{\ast 0}$ and
$D_s^\ast$ respectively. Average values of quark masses yield $r\simeq
0.5$.

In order to analyze the possible NP effects in our observable $\Delta
\theta^{(exp)}$, let us consider the typical situation of a theory
including an extended scalar sector. In the case of multihiggs (MH)
models, the scalar-mediated tree contributions to $C_1$ and $C_2$ can be
neglected, since in general scalar couplings are proportional to the
current quark masses of the involved vertices. On the other hand,
penguin-like diagrams mediated by the new scalars involve vertices which
are proportional to the top quark mass, thus they are potentially
important. Then, while SM amplitudes are dominated by tree contributions
($C_{1,2}^{SM}\gg C_i^{SM}$ for $i=3\dots 6$), in a MH scheme the main
effect of the extended scalar sector on $\alpha_1$ and $\alpha_2$ occurs
through the new contributions to the effective coefficients $a_4$ and
$a_6$. In this way, from Eqs. (\ref{relphase}) and (\ref{acoc}) one gets
\begin{equation}
\Delta \theta^{(exp)} \,
\simeq \, 2 \, \sin \varphi^{MH} \left( \frac{\alpha^{MH}_1}{\alpha^{SM}_1}
- \frac{\alpha^{MH}_2}{\alpha^{SM}_2} \right )\,
\sim  \, - \, 4\, r\,\sin\varphi^{MH}
\frac{|a_6^{MH}|}{a_1}\;.
\label{cpthdm}
\end{equation}
As a first outcome from this expression, it is seen that the ratios
$\alpha_j^{NP}/\alpha_j^{SM}$ do not cancel with each other, consequently
the asymmetry $\Delta \theta^{(exp)}$ is in principle nonzero.

Even if the result in (\ref{cpthdm}) is just an estimate, it can be taken
into account in order to show that new physics effects can be significant
enough to provide an observable signal. According to the analysis
presented in the previous section, this would be achieved if new physics
contributions to $a_6$ reach about 10\% of the SM tree amplitude, and
carry a CP-violating phase of order one (one would obtain in this case an
asymmetry $\Delta \theta^{(exp)}$ of about 10 degrees). Within the SM, the
effective coefficients $|a_1|$ (tree) and $|a_6|$ (penguin) are estimated
to be approximately 1 and $0.06$, respectively \cite{cheng99}. Thus, one
would have important chances of measuring nonstandard physics if new
contributions to $a_6$ carrying large CP-violating phases are comparable
in size to SM ones. It is worth to point out that this level of
contribution of nonstandard physics is indeed suggested by some puzzling
experimental results on penguin-dominated modes, such as the $B\to\eta' K$
branching ratios \cite{etak} and the time-dependent CP asymmetries in
$B^0\to\phi K_S$ \cite{phik}. The experimental values for these
observables are at least $2\sigma$ away from SM expectations, and can be
seen as indications of large new physics effects at the penguin level.

We believe that these experimental observations on penguin-dominated $B$
decay channels already provide a substancial ground to encourage the DP
analysis of $\bddp$ proposed here. On the other hand, we point out that
the room for nonstandard contributions to penguin amplitudes ---and thus
to the phase difference $\Delta \theta^{(exp)}$--- is relatively large,
mainly due to the existing theoretical uncertainties in the evaluation of
SM amplitudes, and to the large number of unknown parameters included in
most scenarios of new physics. To be definite, let us take here as an
example one of the simplest possible extensions of the SM, namely a
two-Higgs-doublet model (THDM) type III. In particular, we consider a
minimal scenario \cite{Cha99} which does not include tree level FCNC, and
the number of new parameters is reduced to four (three Yukawa couplings
parameters plus the charged Higgs mass). In this framework the main new
contributions to $b$ quark decays arise from one-loop diagrams involving a
virtual top quark, while neutral Higgs-mediated diagrams are shown to be
negligible~\cite{Xia01,bsg}. As stated, in this kind of models the largest
new contributions to the amplitudes $a_i$ come through the dipole
coefficient $C_7$, and the allowed space for the new parameters is mainly
constrained by the effects on $B\to X_s\gamma$ decays \cite{bsg}. Taking
into account the bounds in Refs.~\cite{Cha99,bsg}, it is possible to
estimate the allowed values for both the amplitude $a_6$ and the
CP-violating phase $\varphi$. We find that within this model the phase
difference $\Delta \theta^{(exp)}$ can be as large as 3 degrees, which,
according to the analysis Sect.\ III, would be around the limit of
observability for the number of events considered.

The example below should be taken just as an illustration to show the
potentiality of our analysis through a simple manageable case. Clearly,
the inclusion of more degrees of freedom would relax the experimental
bounds on the new model parameters (imposed e.g.\ by the chosen mechanism
to avoid unwanted flavor changing neutral transitions), allowing higher
values for the measurable phase difference $\Delta\theta^{(exp)}$ which
will exceed the observability limits. In addition, other possible
frameworks of nonstandard physics have been shown to provide enhancement
effects on penguin-dominated processes, offering an explanation for the
puzzling time-dependent CP asymmetries in $B^0\to\phi K_S$. Among the most
popular scenarios, recent analyses include R-parity violating
supersymmetry \cite{Dut03}, left-right supersymmetric models \cite{Fra03},
and theories including warped extra dimensions \cite{Bur03}. In all these
models ---which include in general a rather large number of new
parameters---, it has been shown that new physics contributions can be of
the same order as SM penguin amplitudes. In this way, their effects on the
$b\to c\bar c s$ channel could provide an observable signal in the DP
analysis of $\bddp$ decays proposed here.

\section{Summary}

We discuss the possible measurement of nonstandard CP violation in
$\bddp$, exploiting the fact that for these processes the asymmetry
between $B^+$ and $B^-$ decays is expected to be negligibly small in the
Standard Model. The presence of two resonant channels provides the
necessary interference to allow for CP asymmetries in the differential
decay width, even in the limit of vanishing strong rescattering phases.

In order to measure the CP-odd phases entering the interfering
contributions to the total decay amplitude, we propose to use the Dalitz
Plot fit technique. This allows a clean disentanglement of relative
phases, independent of theoretical uncertainties arising from FSI effects.
The expected quality of the experimental measurements has been estimated
by means of a Monte Carlo simulation of the decays, from which we conclude
that the phases can be extracted with a statistical error not larger than
a couple of degrees, provided that the widths of the intermediate $D^{\ast
0}$ and $D_s^\ast$ resonances are at least of the order of a hundred keV.
On the theoretical side, within the framework of generalized factorization
we perform a rough estimation of possible nonstandard CP violation effects
on the interfering amplitudes. We take as an example the typical case of a
multihiggs model, showing that the level of accuracy of the DP fit
measurements can be sufficient to reveal effects of new physics.

Let us finally stress that tree-dominated decays like $\bddp$ are usually
not regarded as good candidates to reveal new physics, since the effects
on branching ratios are not expected to be strong enough to be separated
from the theoretical errors. Our proposal represents a possible way of
detecting these effects by means of CP asymmetries, which can allow the
disentanglement of new physics contributions to penguin-like operators in
a theoretically simple way.

\section*{Acknowledgements}

D.G.D.\ acknowledges financial aid from Fundaci\'on Antorchas (Argentina).
This work has been partially supported by CONICET and ANPCyT (Argentina).


\begin{thebibliography}{999}

\bibitem{Nir94} See e.g.\ Y.\ Nir, H.\ Quinn, in {\em B Decays}, Ed.\ S.\
Stone, World Scientific 1994, p.\ 362.

\bibitem{Nir02} Y.\ Nir, Nucl.\ Phys.\ Proc.\ Suppl.\ {\bf 117}, 111
(2003).

\bibitem{DP} See for example Refs.~\cite{cg-3pi,cg-kpipi}.

\bibitem{cg-3pi} E791 Collab., E.M.\ Aitala {\em et al.}, Phys.\ Rev.\
Lett.\ {\bf 86}, 770 (2001).

\bibitem{cg-kpipi} E791 Collab., E.M.\ Aitala {\em et al.}, Phys.\ Rev.\
Lett.\ {\bf 89}, 121801 (2002).

\bibitem{Bed98} I.\ Bediaga, R.E.\ Blanco, C.\ G\"obel, R.\ Mendez
Galain, Phys.\ Rev.\ Lett.\ {\bf 81}, 4067 (1998); R.E.\ Blanco, C.\
G\"obel, R.\ Mendez Galain, Phys.\ Rev.\ Lett.\ {\bf 86}, 2720 (2001).

\bibitem{D-decays} See for example Refs.~\cite{cg-3pi,cg-kpipi}.
A complete list can be found in \cite{PDG}.

\bibitem{PDG} Particle Data Group, K.\ Hagiwara {\em et al.}, Phys.\
Rev.\ D {\bf 66}, 010001 (2002).

\bibitem{Atw94} This idea has been already exploited by several authors.
See e.g.\ D.\ Atwood, A.\ Soni, Z.\ Phys. C {\bf 64} 221 (1994); {\em
ibid.} Phys.\ Rev.\ Lett.\ {\bf 74} 220 (1995); D.\ Atwood, G.\ Eilam, M.\
Gronau, A.\ Soni, Phys.\ Lett.\ B {\bf 341} 372 (1994); R.\ Enomoto, Y.\
Okada, Y.\ Shimizu, Phys.\ Lett.\ B {\bf 433} 109 (1998).

\bibitem{Buch96} G.\ Buchalla, A.J.\ Buras, M.\ Lautenbacher, Rev.\
Mod.\ Phys.\ {bf 68}, 1125 (1996).

\bibitem{Gir01} A.K.\ Giri, R.\ Mohanta, M.P.\ Khanna, Eur.\ Phys.\
J.\ C {\bf 22}, 115 (2001).

\bibitem{Ger91} J.-M.\ G\'erard, W.-S\ Hou, Phys.\ Rev.\ D {\bf 43},
2909 (1991).

\bibitem{Bed97} I.\ Bediaga, C.\ G\"obel, R.\ Mendez Galain, Phys.\ Rev.\
Lett.\ {\bf 78}, 22 (1997).

\bibitem{Cin01} CLEO Collab., S.\ Ahmed {\em et al.}, Phys.\ Rev.\ Lett.\
{\bf 87}, 251801 (2001).

\bibitem{bau87} M.\ Bauer, B.\ Stech, M.\ Wirbel, Z.\ Phys.\ C {\bf
34}, 103 (1987).

\bibitem{Kim02} Z.\ Luo, J.L.\ Rosner, Phys.\ Rev.\ D {\bf 64}, 094001 (2001);
C.S.\ Kim, Y.\ Kwon, J.\ Lee, W.\ Namgung, Phys.\ Rev.\ D {\bf 65}, 097503
(2002).

\bibitem{AliGre98} A.\ Ali, C.\ Greub, Phys.\ Rev.\ D {\bf 57}, 2996 (1998).

\bibitem{Xia01} Z.\ Xiao, C.S.\ Li, K.-T.\ Chao, Phys.\ Rev.\ D {\bf
63}, 074005 (2001); Z.\ Xiao, K.-T.\ Chao, C.S.\ Li, Phys.\ Rev.\ D {\bf
65} 114021 (2002).

\bibitem{Bur98} A.J.\ Buras, R.\ Fleischer, Adv.\ Ser.\ Direct.\ High
Energy Phys.\ {\bf 15}, 65 (1998).

\bibitem{Ali98} A.\ Ali, G.\ Kramer, C.-D.\ L\"u, Phys.\ Rev.\ D
{\bf 58} 094009 (1998).

\bibitem{Che99} Y.-H.\ Chen, H.-Y.\ Cheng, B.\ Tseng, K.-C.\ Yang, Phys.\
Rev.\ D {\bf 60}, 094014 (1999).

\bibitem{Cheli99} H.-Y.\ Cheng, H.-n.\ Li, K.-C.\ Yang, Phys.\
Rev.\ D {\bf 60}, 094005 (1999).

\bibitem{cheng99} H.-Y.\ Cheng, K.-C.\ Yang, Phys.\ Rev.\ D {\bf 59},
092004 (1999).

\bibitem{Kim01} C.S.\ Kim, Y.\ Kwon, J.\ Lee, W.\ Namgung, Phys.\ Rev.\ D
{\bf 63}, 094506 (2001).

\bibitem{etak} CLEO Collab., S.J.\ Richichi {\em et al.}, Phys.\ Rev.\
Lett.\ {\bf 85}, 520 (2000); Belle Collab., K.\ Abe {\em et al.}, Phys.\
Lett.\ B {\bf 517}, 309 (2001); BaBar Collab., B.\ Aubert {\em et al.},
{\tt hep-ex/0303046}.

\bibitem{phik} BaBar Collab., B.\ Aubert {\em et al.}, {\tt
hep-ex/0207070} (ICHEP 02); Belle Collab., K.\ Abe {\em et al.},
Phys.\ Rev.\ D {\bf 67}, 031102(R) (2003).

\bibitem{Cha99} D.\ Bowser-Chao, K.-m.\ Cheung, W.-Y.\ Keung, Phys.\ Rev.\ D
{\bf 59}, 115006 (1999)

\bibitem{bsg} Z.J.\ Xiao, C.S.\ Li, K.T.\ Chao, Phys.\ Rev.\ D {\bf 62},
094008 (2000).

\bibitem{Dut03} B.\ Dutta, C.S.\ Kim, S.\ Oh, Phys.\ Lett.\ B {\bf 535}, 249;
Phys.\ Rev.\ Lett.\ {\bf 90}, 011801 (2003); A.\ Kundu, T.\ Mitra, Phys.\
Rev.\ D {\bf 67}, 116005 (2003).

\bibitem{Fra03} M.\ Frank, Phys.\ Rev.\ D {\bf 68}, 035011 (2003).

\bibitem{Bur03} G.\ Burdman, {\tt hep-ph/0310144}.

\end{thebibliography}
\end{document}